\documentclass[12pt,a4paper]{JHEP}
\usepackage{amsmath,amssymb,amsfonts,graphicx,cite,multirow}
\graphicspath{{graphics/}}
\makeatletter
\ifx\input@path\@undefined
\def\input@path{{graphics/}}
\else
\g@addto@macro\input@path{{graphics/}}
\fi
\makeatother
\newcommand{\program}[1]{\textsf{#1}}

\newcommand{\figref}[1]{Fig.~\ref{#1}}
\newcommand{\secref}[1]{Sec.~\ref{#1}}

\newcommand{\Tr}{\mbox{Tr}}
\newcommand{\TR}{\ensuremath{T_R}}
\newcommand{\cf}{\ensuremath{C_F}}
\newcommand{\as}{\ensuremath{\alpha_s}}
\newcommand{\Nc}{\ensuremath{N_c}}
\newcommand{\Np}{\ensuremath{n}}

\newcommand{\Ng}{\ensuremath{N_g}}

\newcommand{\qbar}{\ensuremath{\overline{q}}}
\newcommand{\qqbar}{\ensuremath{q\qbar }}

\preprint{DESY-11-255\\LU-TP-11-48\\MCnet-11-27}

\title{Subleading $\Nc$ improved Parton Showers}

\author{Simon Pl\"atzer ${}^{a}$ and Malin Sj\"odahl ${}^{b}$}

\institute{${}^{a}$ DESY,\\Notkestrasse 85, D-22607 Hamburg, Germany\\\vspace*{1ex}
  ${}^{b}$ Dept. of Astronomy and Theoretical Physics, Lund University,\\
  S\"olvegatan 14A, SE-223 62 Lund, Sweden}

\date{\today}

\abstract{We describe an algorithm for improving subsequent parton
  shower emissions by full SU(3) color correlations in the framework
  of a dipole-type shower.  As a proof of concept, we present
  results from the first implementation of such an algorithm for a final
  state shower.  The corrections are found to be small for event
  shapes and jet rates but can be more significant for tailored
  observables.}


\begin{document}

\sloppy
\section{Introduction}
\label{sections:introduction}

Parton showers and event generators are indispensable tools for predicting
and understanding collider results\cite{Sjostrand:2007gs,Bahr:2008pv,Gleisberg:2003xi}. 
Considering their importance for interpreting 
LHC results, it is essential to have a good understanding of 
their approximations and limitations.

Significant progress has been made over the last years in the areas of
matrix element merging at leading order
\cite{Lonnblad:2001iq,Krauss:2002up,Hoche:2006ph,
  Lavesson:2007uu,Hoeche:2009rj,Hamilton:2009ne}, and matching at next
to leading order
\cite{Dobbs:2001dq,Frixione:2002ik,
  Nason:2004rx,Nagy:2005aa,Frixione:2010ra,Platzer:2011bc}, as well as
improvements to shower algorithms using dipole-type evolution
\cite{Nagy:2006kb,Winter:2007ye,Dinsdale:2007mf,Schumann:2007mg,Platzer:2009jq},
partly generalizing ideas presented originally in \cite{Gustafson:1987rq}.
There has also been theoretical progress in the direction of treating full
SU(3) at the level of shower evolution \cite{Nagy:2007ty},
and at the level of dealing with the color space 
\cite{Nagy:2007ty,Sjodahl:2009wx,Sjodahl:2008fz}.

In dipole-type parton showers, only large $\Nc$ color connected
partons\footnote{`Color connected' here means that two partons share a
  common color line as used in the double line notation. Note that the
  double line notation can be extended beyond the large $\Nc$ limit when
  including an appropriate singlet contribution for the color
  structure of the each gluon propagator.} can radiate coherently. The,
typically $1/\Nc^2$ suppressed, coherent emission from other non-color
neighboring partons is neglected. However, as the number of
perturbatively emitted partons $\Np$ increases, the number of possible
contributions from non color neighbors grows roughly like $\Np^2/2$,
while the number of possible coherent emissions from color neighbors
grows only like $\Np$. With the higher multiplicities at high energy
hadron colliders, these color suppressed contributions may thus become
important, though this counting does not include the dynamics of
multiple parton emission. A treatment of these color suppressed
contributions within a full-fledged shower framework is thus
necessary.

This paper reports on a first step in the direction of incorporating
color suppressed contributions by iterative use of color matrix
element corrections.  We outline an algorithm to calculate these color
matrix element corrections to improve shower emissions by the full
SU(3) color correlations and present a proof of concept implementation
based on the shower implementation described in
\cite{Platzer:2009jq,Platzer:2011bc}. The first results are presented for
$e^+e^-\to \text{jets}$.

For the implementation, a \program{C++} color algebra package, 
\program{ColorFull} \cite{Sjodahl:Colorfull}, has been developed 
for keeping track of the color structure and aiding
calculation of high multiplicity matrix elements squared and color
correlated matrix elements. It has also been necessary to consider the
treatment of negative weights in the context of partons showers, as
there are negative contributions to the radiation probability from
color suppressed terms \cite{Platzer:2011dq}.

This paper is organized as follows: in
\secref{sections:Dipole} we recapitulate the dipole factorization
scheme and the dipole shower evolution.  In \secref{sections:cmes} we
define an algorithm for the inclusion of full color correlations for
subsequent parton shower emissions.  The details of the color space
treatment are given in \secref{sections:colorbasis}.  In
\secref{sections:results} we present numerical results from the
implementation of a parton shower improved by full color
correlations. \secref{sections:conclusions} draws some conclusions
and gives an outlook on future developments.

\section{Dipole Factorization and Dipole Shower Evolution}
\label{sections:Dipole}

Dipole factorization, 
\cite{Catani:1996jh,Catani:1996vz},
states that the behavior of QCD tree-level matrix elements squared in
any singly unresolved limit involving two partons $i,j$ ({\it i.e.}
whenever $i$ and $j$ become collinear or one of them soft), can be
cast into the form
\begin{multline}
  \label{eqs:dipolefactorization}
  |{\cal M}_{n+1}(...,p_i,...,p_j,...,p_k,...)|^2 \approx\\
  \sum_{k\ne i,j} \frac{1}{2 p_i\cdot p_j}
  \langle {\cal M}_n(...,p_{\tilde{ij}},...,p_{\tilde{k}},...) |
          {\mathbf V}_{ij,k}(p_i,p_j,p_k)| {\cal M}_n(...,p_{\tilde{ij}},...,p_{\tilde{k}},...)\rangle \ ,
\end{multline}
where $|{\cal M}_{n}\rangle$ -- which is a vector in the space of
helicity and color configurations -- denotes the amplitude for an
$n$-parton final state.  Here an emitter $\tilde{ij}$ undergoes
splitting to two partons $i$ and $j$ in the presence of a spectator
$\tilde{k}$ which absorbs the longitudinal recoil of the splitting,
$\tilde{k}\to k$, such as to keep the momenta both before and after
emission on their mass shell while maintaining exact energy-momentum
conservation.  Considering gluon emission only, this can be
interpreted as a dipole $\tilde{ij},\tilde{k}$ emitting a gluon of
momentum $p_j$.  (However, $g\to \bar{q}q$ splittings fit into this
framework without conceptional changes).

In general, the insertion operator ${\mathbf V}$ contains both color
correlations stemming from soft gluon emissions and spin correlations
originating in the collinear splitting of a gluon. We shall limit
ourselves to massless quarks and color correlations only, using the
spin averaged dipole splitting functions presented in
\cite{Catani:1996vz}.  In this case
\begin{equation}
{\mathbf V}_{ij,k}(p_i,p_j,p_k) = 
-8\pi\alpha_s V_{ij,k}(p_i,p_j,p_k) 
\frac{{\mathbf T}_{\tilde{ij}}\cdot {\mathbf T}_k}{{\mathbf T}_{\tilde{ij}}^2}
\end{equation}
in terms of the color charge operators ${\mathbf T}_i$\footnote{We use
  the conventions on the color charge algebra as given in
  \cite{Catani:1996vz}. Note that the Casimir operators ${\mathbf
    T}_i^2=C_F$ if $i$ is an (anti-)quark, and ${\mathbf T}_i^2=C_A$
  if $i$ is a gluon are included in the definition of the splitting
  kernels $V$; thus the color correlations are normalized
  accordingly. This convention is more transparent when comparing to
  parton showers in the large $\Nc$ limit to be discussed below.}.
This notation is independent of the basis considered for color space,
though we shall stick to one particular choice of basis, to be
discussed in \secref{sections:colorbasis}.

\FIGURE[t]{
\includegraphics[scale=0.6]{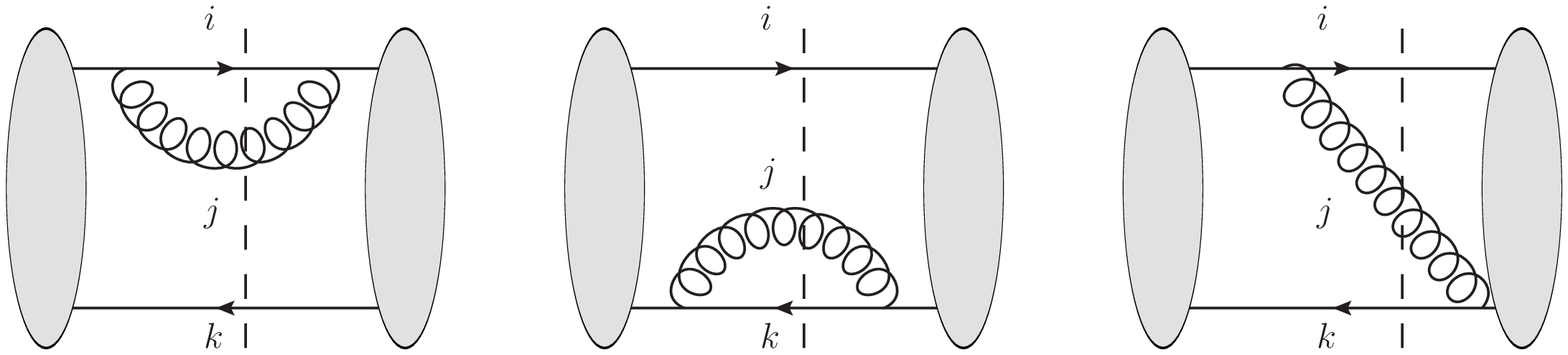}
\caption{\label{fig:dipoles}Sample diagrams contributing to the dipole
  kernels ${\mathbf V}_{ij,k}$ and ${\mathbf V}_{kj,i}$ in the
  singular limits for gluon emission off a quark or antiquark. Similar
  diagrams are present for gluon splittings.}  } The kernel $V_{ij,k}$
only contains a collinear enhancement with respect to the splitting
$\tilde{ij}\to i,j$.  It is constructed by rearranging singular
contributions of sets of diagrams as depicted in \figref{fig:dipoles}
while making use of the fact that the amplitude is a color singlet,

\begin{equation}
{\mathbf T}_{\tilde{ij}}^2 = - \sum_{k\ne \tilde{ij}} {\mathbf T}_{\tilde{ij}}\cdot {\mathbf T}_k \ .
\end{equation}
For gluon emission off a final-final dipole configuration, 
the case considered in the present paper, the kernels read
\begin{eqnarray}
V_{qg,k}(p_i,p_j,p_k) &=& \cf \left( \frac{2 (1-z)}{(1-z)^2+p_\perp^2/s_{ijk}} - (1+z) \right)\\
V_{gg,k}(p_i,p_j,p_k) &=& 2C_A\left( \frac{1-z}{(1-z)^2+p_\perp^2/s_{ijk}} + 
\frac{z}{z^2+p_\perp^2/s_{ijk}} - 2 + z(1-z) \right)
\end{eqnarray}
where we have introduced the relative scalar transverse momentum
$p_\perp$ and longitudinal momentum fraction $z$ of parton
$i$ as given by the Sudakov decomposition

\begin{eqnarray}
p_i & = & z p_{\tilde{ij}} +\frac{p_\perp^2}{z s_{ijk}}p_{\tilde{k}} + k_\perp \\
p_j & = & (1-z) p_{\tilde{ij}} +\frac{p_\perp^2}{(1-z) s_{ijk}}p_{\tilde{k}} - k_\perp \\
p_k & = & \left(1-\frac{p_\perp^2}{z(1-z)s_{ijk}}\right)p_{\tilde{k}} \;,
\end{eqnarray}
with $p_{\tilde{ij}}^2=p_{\tilde{k}}^2=0$, a space like transverse
momentum $k_\perp$ with $k_\perp^2=-p_\perp^2$ and $k_\perp\cdot
p_{\tilde{ij}}= k_\perp\cdot p_{\tilde{k}}= 0$. With this
parametrization we also have
$s_{ijk}=(p_i+p_j+p_k)^2=(p_{\tilde{ij}}+p_{\tilde{k}})^2$.

Reinterpreting the dipole factorization in terms of emitting dipoles,
parton cascades can be based on it and, after slight modifications to
initial state radiation, have been shown to exhibit the proper
features in terms of coherent gluon emission and logarithmic accuracy
\cite{Platzer:2009jq}. 
The present work is an extension to the
implementation of a coherent dipole evolution, details of which have
been reported in \cite{Platzer:2011bc}. Besides using spin averaged
dipole kernels -- like any other parton shower implementation available so far --
the Monte Carlo implementation in \cite{Platzer:2011bc}
treats the color correlation operator for the shower in the large $\Nc$ limit
{\it i.e.} when using the double line notation of color

\begin{equation}
  -\frac{{\mathbf T}_{\tilde{ij}}\cdot {\mathbf T}_k}{{\mathbf T}_{\tilde{ij}}^2}\to
  \frac{1}{1+\delta_{\tilde{ij}}}\delta(\tilde{ij}, k\text{ color connected})\ ,
  \label{eq:InfCol}
\end{equation}
where $\delta_{\tilde{ij}}=1$ if $\tilde{ij}$ is a gluon, and vanishes otherwise.
Here the initial assignment and further evolution of large $\Nc$
color charge flow is extensively discussed in {\it e.g.}
\cite{Platzer:2011bc}. 
Let us only recall here that to each of the two
emitter-spectator partitions of a large $\Nc$ color connected pair of
partons we connect an emission probability related to the dipole
kernels $V_{ij,k}$ as
\begin{equation}
  \label{eq:Pleading}
{\rm d}P_{ij,k}(p_\perp^2,z;p_{\tilde{ij}},p_{\tilde{k}}) = 
\frac{\alpha_s}{2\pi}\frac{{\rm d}p_\perp^2}{p_\perp^2} {\rm d}z
{\cal J}(p_\perp^2,z;p_{\tilde{ij}},p_{\tilde{k}}) V_{ij,k}(p_\perp^2,z;p_{\tilde{ij}},p_{\tilde{k}}) \;.
\end{equation}
Here, $\cal{J}$ represents the Jacobian for changing variables from
$p_i,p_j,p_k$ to $p_{\tilde{ij}}$, $p_{\tilde{k}}$, $p_\perp$, $z$ and an azimuthal orientation,
which has been integrated over. For a
final-final dipole configuration, ${\cal
  J}(p_\perp^2,z;p_{\tilde{ij}},p_{\tilde{k}})=(1-p_\perp^2/z(1-z)s_{ijk})$.

In this paper we report on generalizing this picture from large $\Nc$
color connected dipoles to all pairs of partons, taking into account
the exact evaluation of the color correlations. Thus, the
notion of dipole chains and their splittings does not apply
anymore. Instead every pair of partons has to be considered competing for
the next emission. Details of the generalized algorithm are given
in \secref{sections:cmes} below.

\section{Color Matrix Element Corrections}
\label{sections:cmes}

Following the arguments of how a parton cascade is constructed from
the dipole factorization formula \eqref{eqs:dipolefactorization}, we
may recast \eqref{eq:Pleading} to include the exact color
correlations by assigning splitting rates
\begin{multline}
{\rm d}P_{ij,k}(p_\perp^2,z;p_{\tilde{ij}},p_{\tilde{k}}) = \\
\frac{\alpha_s}{2\pi}\frac{{\rm d}p_\perp^2}{p_\perp^2} {\rm d}z
{\cal J}(p_\perp^2,z;p_{\tilde{ij}},p_{\tilde{k}}) V_{ij,k}(p_\perp^2,z;p_{\tilde{ij}},p_{\tilde{k}})
\times \frac{-1}{{\mathbf T}_{\tilde{ij}}^2} \frac{\langle
  {\cal M}_n|{\mathbf T}_{\tilde{ij}}\cdot {\mathbf T}_k |{\cal M}_n\rangle
}{|{\cal M}_n|^2}
\label{eq:dP}
\end{multline}
to {\it each  pair} $\tilde{ij},\tilde{k}$ of partons to generate an emission 
$j$ off an $n$-parton system.  In the large $\Nc$ limit the factor after the
times sign reduces to \eqref{eq:InfCol} for color connected pairs of partons and
zero otherwise. Here, by allowing any pair of partons to radiate,
it accounts for correcting the emission to include the exact color
correlations. Owing to the similarity to so-called matrix element
corrections present in parton showers,
\cite{Seymour:1994df,Norrbin:2000uu}, we refer to this improvement as
     {\it color matrix element corrections}.

At the level of the hard process, from which the shower evolution is
starting, the calculation of $|{\cal M}_n\rangle$ is
straightforward, but once the first emission is performed, the issue of how to 
describe the evolved state has to be addressed.
By picking a particular basis for the color structures, we can map
$|{\cal M}_n\rangle$ to a complex vector ${\cal M}_n\in {\mathbb
  C}^{d_n}$, where $d_n$ is the dimensionality of the basis for $n$
partons,
\begin{equation}
|{\cal M}_n\rangle = \sum_{\alpha=1}^{d_n} c_{n,\alpha} |\alpha_n\rangle
\quad \leftrightarrow \quad {\cal M}_n = (c_{n,1},...,c_{n,d_n})^T \ .
\end{equation}
Then we have
\begin{equation}
\label{eqs:m2}
|{\cal M}_n|^2 = {\cal M}_n^\dagger S_n {\cal M}_n =
\Tr \left( S_n\times {\cal M}_n{\cal M}_n^\dagger \right) 
\end{equation}
with $S_n$ being the scalar product matrix, 
$(S_n)_{\alpha\beta}=\langle\alpha_n|\beta_n\rangle$,
for color basis vectors $|\alpha_\Np \rangle$ and $|\beta_\Np \rangle$.
Moreover, the color correlated matrix element for emission from 
$\tilde{ij}$ and $\tilde{k}$
can be written as 
\begin{equation}
\label{eqs:ccm2}
\langle {\cal M}_n|{\mathbf T}_{\tilde{ij}}\cdot {\mathbf T}_{\tilde{k}}|{\cal M}_n\rangle = 
\Tr \left( S_{n+1}\times T_{\tilde{k},n} {\cal M}_n{\cal M}_n^\dagger T_{\tilde{ij},n}^\dagger \right)
\end{equation}
in terms of matrix representations of ${\mathbf
  T}_{\tilde{ij}},{\mathbf T}_{\tilde{k}}\in {\mathbb
  C}^{d_{n+1},d_n}$.  These matrices describe the color space effect
of emission from the partons $\tilde{ij}$ and $\tilde{k}$ by mapping a
basis tensor in $n$-parton space to a linear combination of basis
tensors in $n+1$-parton space.  In the "trace basis", which is used
here, (c.f. \secref{sections:colorbasis}) the ${\mathbf
  T}_{\tilde{ij}}$ matrices are very sparse
\cite{Nagy:2007ty,Sjodahl:2009wx}.  
The scalar product matrices,
$S_n,S_{n+1}$, are however dense, and
the calculations of these matrices is a key ingredient for running a
parton shower improved by color matrix element corrections.

As we want to keep the full color structure the evolution is (analogous
to \cite{Nagy:2007ty}) done keeping the amplitude information, using a
matrix $M_n$, which initially -- for the hard process the shower
starts with -- is given by $M_n={\cal M}_n{\cal M}_n^\dagger$.
After having performed an emission with momentum $p_j$ off an
$n$-parton configuration, this quantity for the $n+1$ parton
configuration can then be obtained from the spin-averaged dipole
kernel as
\begin{equation}
\label{eqs:nextamplitude}
M_{n+1} =
-\sum_{i\ne j}\sum_{k\ne i,j} \frac{4\pi\alpha_s}{p_i\cdot p_j} \frac{V_{ij,k}(p_i,p_j,p_k)}{{\mathbf T}_{\tilde{ij}}^2}
\ T_{\tilde{k},n}M_n T_{\tilde{ij},n}^\dagger \ .
\end{equation}
Matrix elements squared and color correlated matrix elements for each
subsequent color matrix element correction are then calculated as in
\eqref{eqs:m2} and \eqref{eqs:ccm2} upon replacing ${\cal M}_n{\cal
  M}_n^\dagger\to M_n$.  One property of the color matrix element
correction weight is that it is not necessarily positive definite for
subleading-$\Nc$ contributions.  The authors have shown in
\cite{Platzer:2011dq}, that this does not pose a problem for a Monte
Carlo implementation. In particular, from the very definition of the
dipole factorization as an approximation to a tree-level matrix
element squared, the sum of the splitting probabilities including all
pairs $i,k$ is strictly positive definite. Thus we can immediately
apply the interleaved competition/veto algorithm \cite{Platzer:2011dq}
to generate kinematic variables with the desired density.

More precisely we select a set of candidate emissions $\tilde{ij},\tilde{k}\to
i,j,k$ at scales $p_{\perp,\tilde{ij},k}$ (when starting from a scale $Q_\perp^2$
associated with the dipole $\tilde{ij},\tilde{k}$) according to the Sudakov form
factor
\begin{equation}
-\ln \Delta_{ij,k}(p_{\perp,ij,k}^2|Q_\perp^2) =
\frac{\alpha_s}{2\pi}\int_{p_{\perp,ij,k}^2}^{Q_\perp^2} \frac{{\rm d}q_\perp^2}{q_\perp^2}
\int_{z_-(q_\perp^2,Q_\perp^2)}^{z_+(q_\perp^2,Q_\perp^2)}{\rm d}z\ 
{\cal P}_{ij,k}(q_\perp^2,z;p_{\tilde{ij}},p_{\tilde{k}}) \ ,
\end{equation}
if ${\cal P}_{ij,k}$ is positive. Here, in accordance with \eqref{eq:dP},
\begin{equation}
{\cal P}_{ij,k}(p_\perp^2,z;p_{\tilde{ij}},p_{\tilde{k}})
={\cal J}(p_\perp^2,z;p_{\tilde{ij}},p_{\tilde{k}}) V_{ij,k}(p_\perp^2,z;p_{\tilde{ij}},p_{\tilde{k}})
\times \frac{-1}{{\mathbf T}_{\tilde{ij}}^2} \frac{\langle
  {\cal M}_n|{\mathbf T}_{\tilde{ij}}\cdot {\mathbf T}_k |{\cal M}_n\rangle
}{|{\cal M}_n|^2}
\end{equation}
and $z_\pm(p_\perp^2,Q_\perp^2)=(1\pm
\sqrt{1-p_\perp^2/Q_\perp^2})/2$. Amongst the candidate splittings of
$\tilde{ij},\tilde{k}$, we pick the one with largest 
$p_{\perp,\tilde{ij},k}^2=p_\perp^2$ and accept the
corresponding splitting with probability
\begin{equation}
\frac{\sum_{\tilde{ij}}\sum_{\tilde{k}\ne \tilde{ij}} {\cal P}_{ij,k}(p_\perp^2,z;p_{\tilde{ij}},p_{\tilde{k}})}
{\sum_{\tilde{ij}}\sum_{\tilde{k}\ne \tilde{ij}}{\cal P}_{ij,k}(p_\perp^2,z;p_{\tilde{ij}},p_{\tilde{k}})
\theta({\cal P}_{ij,k}(p_\perp^2,z;p_{\tilde{ij}},p_{\tilde{k}}))}\;.
\end{equation}
Upon rejection, the selection is repeated, setting $Q_\perp=p_\perp$,
until the next $p_\perp$ is accepted or eventually $p_\perp$ is found
at the infrared cutoff $\mu={\cal O}(1\ {\rm GeV})$.  Upon accepting
the candidate splitting with scale $p_\perp$, the dipole
$\tilde{ij},\tilde{k}$ is chosen to define the Sudakov decomposition
of the emission kinematics. The generated emission is then inserted
into the event record, and $M_{n+1}$ is 
determined using the generated kinematics according to 
\eqref{eqs:nextamplitude}.

\section{Color Basis Treatment}
\label{sections:colorbasis}

Throughout this paper we use the basis 
obtained by first connecting all 
$q\qbar$-pairs in all possible ways (for one $\qqbar$-pair only one way), 
and then attaching gluons to the quark-lines in 
all possible ways ($\Ng!$ ways for one $\qqbar$-pair), 
see {Fig.~\ref{fig:Basis}a}. This basis is sufficient as 
long as only leading order QCD processes are considered.
In the presence of (QCD) virtual corrections, or processes mediated by
electroweak (or other color singlet) exchanges, we would also have to
consider basis vectors obtained from direct products of open and closed
quark-lines, as depicted in \figref{fig:Basis}.

\FIGURE[t]{
  \includegraphics{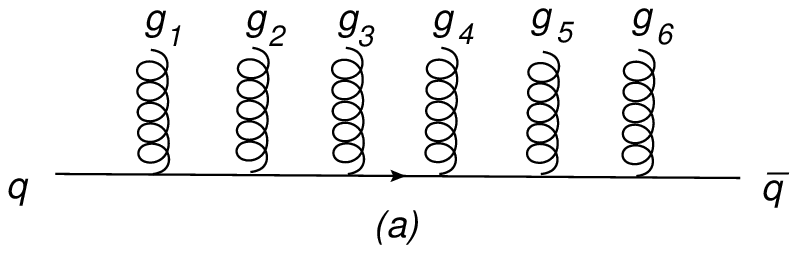}
  \includegraphics{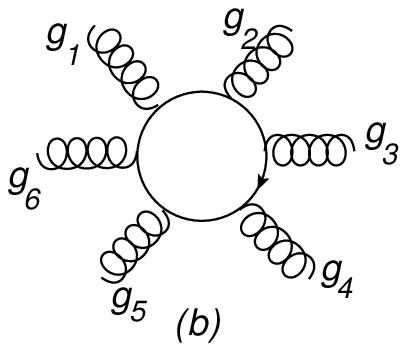}
  \caption{\label{fig:Basis} Examples of color structures.  (a): An
    open quark line with 6 gluons corresponding to the color structure
    $t^{g_1}_{q q_1}t^{g_2}_{q_1 q_2}t^{g_3}_{q_2 q_3}t^{g_4}_{q_3
      q_4}t^{g_5}_{q_4 q_5}t^{g_6}_{q_5 \bar{q}}$. This is the only type of
    color structure needed for 
    $e^+e^-\rightarrow \qqbar \rightarrow\qqbar +\Ng$ gluons at leading order. 
    (b) A closed quark
    line corresponding to the color structure $t^{g_1}_{q_6
      q_1}t^{g_2}_{q_1 q_2}t^{g_3}_{q_2 q_3}t^{g_4}_{q_3
      q_4}t^{g_5}_{q_4 q_5}t^{g_6}_{q_5 q_6}= \mbox{Tr}\left[
      t^{g_1}t^{g_2}t^{g_3}t^{g_4}t^{g_5}t^{g_6} \right]$. }} 

Due to the relation 
\begin{equation}
if_{abc}=\frac{1}{\TR}(\Tr[t^a t^b t^c] - \Tr[t^b t^a t^c])=
\frac{1}{\TR}(t^a_{q_1 q_2} t^b_{q_2 q_3}t^c_{q_3 q_1} -t^b_{q_1 q_2} t^a_{q_2 q_3}t^c_{q_3 q_1})
\label{eq:f}
\end{equation}
where $\TR=\Tr[t^at^a]$ (no sum) is taken to be $1/2$,
the effect of gluon emission and gluon exchange is trivial in this
basis \cite{Nagy:2007ty,Sjodahl:2009wx}. Graphically the emission process 
can be exemplified as in \figref{fig:Emission}.  
Using the notation
\begin{equation}
  |\{ q\, g_1\,g_2...g_m \,\qbar\} \rangle=
  t^{g_1}_{q,q_1}t^{g_2}_{q_1,q_2}...t^{g_m}_{q_{m-1},\qbar}
\end{equation}
we may in general write for the emission of gluon $g_{n+1}$ from the gluon at 
place $i$
\begin{equation}
  |\{ q\, g_1...\tilde{g}_i...g_m \,\qbar\} \rangle
  \rightarrow 
  |\{ q\, g_1...g_i\,g_{m+1}\,...g_m\, \qbar\} \rangle
  - |\{ q\, g_1...g_{m+1}\,g_i...g_m\, \qbar\}\rangle
  \label{eq:Emission}
\end{equation}
where the overall sign depends on the sign convention for the
triple gluon vertex, and thus has to be matched with the convention for
the momentum dependent part. Here we choose the triple gluon vertex factor to be
$i f_{\tilde{g},g,g_{m+1}}$ 
where $\tilde{g}$ denotes the emitters color index before
emission, $g$ denotes the emitters color after emission, and $g_{m+1}$ denotes
the color index of the emitted gluon.  
For quarks and anti-quarks the effects of gluon emission are similarly
\begin{eqnarray}
  |\{ q\, g_1...g_m \,\qbar\}\rangle
  &\rightarrow& \;\;\;|\{ q\, g_{m+1}\,g_1...g_m \,\qbar\}\rangle \;\mbox{    for }q\\
  |\{ q\, g_1...g_m \,\qbar\}\rangle
  &\rightarrow& 
  -|\{ q\,g_1...g_m \, g_{m+1}\,\qbar\} \rangle \;\mbox{   for \qbar}.
\end{eqnarray}
\FIGURE[t]{
  \includegraphics{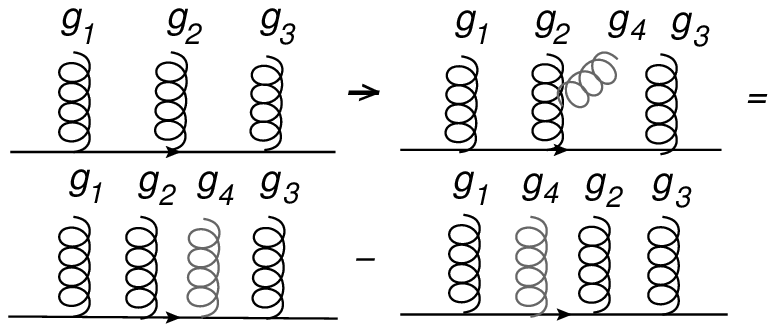}
  \caption{\label{fig:Emission}
The graphical representation of \eqref{eq:Emission} for $i=2$.
}}

The disadvantages of this type of basis -- obtained by connecting
partons in all possible ways -- is that it is orthogonal only in the
$\Nc \rightarrow \infty$ limit, and that it is overcomplete for finite
$\Nc$. For $\Nc=3$ the lowest number of partons for which this kind of
basis is overcomplete is four gluons (incoming plus outgoing).

We note that there are two computational steps which naively scale as 
$(\Ng!)^2$. The first step is the step of calculating all the
scalar products between the $\Ng!$ basis vectors. The second is the step of
identifying the new basis vectors when a given parton in a given 
basis vector has emitted a new gluon;
the number of basis vectors in the initial basis scales as $(\Ng-1)!$ 
(starting from $\Ng-1$ gluons),
and the number of basis vectors in the final basis as $\Ng!$. 
This would give an overall scaling behavior of 
$(\Ng-1)!(\Ng+1)\Ng!\sim (\Ng!)^2$, 
as there are $(\Ng+1)$ possible emitters.

However, enumerating the basis states in a unique way, it is possible
to {\it calculate} the numbers of the new basis vectors when a gluon
has been emitted from a given parton in a given basis vector. As there
is no need to compare the result after emission to all basis vectors
in the $\Ng$-basis, this reduces the computational effort to scale
rather as $(\Ng-1)!(\Ng+1)$ in the step of identifying all new color
states when starting in $(\Ng-1)!$ possible basis vectors and emitting
from $\Ng+1$ possible emitters.  This is clearly much better than the
$(\Ng!)^2$ scaling for calculating the scalar products, which thus is
the major bottleneck for sufficiently many partons. Fortunately these
calculations need only be performed once, and can then be stored 
numerically for use in the color matrix element corrections.

The scalar product matrices have been checked to agree with scalar product 
matrices calculated by the Mathematica code published along with 
\cite{Sjodahl:2008fz} for up to 4 gluons, and against a new
(yet unpublished) Mathematica package.
Similarly the matrices representing 
${\mathbf T}_{\tilde{ij}}\cdot {\mathbf T}_{\tilde{k}}$
have been checked to agree for up to 3 gluons compared to 
the old Mathematica code, and for up to 4 gluons compared to the new.

\section{Results}
\label{sections:results}

We here discuss first results from a subleading
$\Nc$ improved final state shower, originating from $e^+e^-\to q
\bar{q}$ at $91\ {\rm GeV}$ center of mass energy. We consider gluon
emission only, as there is no soft enhancement present for a $g\to
q\bar{q}$ splitting.
The strong coupling constant is taken to be fixed, $\as=0.112$.
The subsequent gluon emissions are performed as described in 
\secref{sections:cmes} using one of three options for the color structure:

\begin{enumerate}
\item {{\bf Full} color structure; the full SU(3) color structure 
  with splitting kernels as in \eqref{eq:dP}
  is used with emissions generated as in \secref{sections:cmes}.}
\item {{\bf Shower} color; the color matrix element correction is
  evaluated in the large $\Nc$ limit, though the color factor $\cf$
  entering $q\to qg$ splittings is kept at its exact value
  $\cf=4/3$. This resembles current parton shower implementations.}
\item {{\bf Strict large} $\Nc$, all $\Nc$ suppressed terms are dropped,
  implying $\cf=3/2$.}
\end{enumerate}
Before turning to the numerical results it is instructive to study the
expected effects already at an analytical level, and to
clarify how the shower limit maps to the standard shower
implementation. We therefore consider the correction weights for up to
three emissions, labelling each intermediate state with $\Ng=n-2$
additional gluons as $q_1\bar{q}_2g_3...g_n$. The gluon $g_n$,
is always considered to be the one emitted in the last
transition, and the weight associated to a dipole $i,k$ in an ensemble
with $n$ partons is denoted by
\begin{equation}
\frac{1}{{\mathbf T}_i^2}\frac{4\pi \alpha_s}{p_i\cdot p_n} V_{in,k}(p_i,p_n,p_k) \equiv V^n_{ik} \ .
\end{equation}
The color matrix element correction for emission off a dipole $i,k$ in
an $n$-parton ensemble is denoted 
\begin{equation}
-\frac{1}{{\mathbf T}_i^2} \frac{\langle {\cal M}_n|{\mathbf T}_i\cdot {\mathbf T}_k|{\cal M}_n\rangle}{|{\cal M}_n|^2}
= -\frac{1}{{\mathbf T}_i^2}
\frac{\Tr \left( S_{n+1}\times T_{i,n} M_n T_{k,n}^\dagger \right)}{
\Tr \left( S_{n}\times M_n\right)
}
\equiv w^n_{ik}
\end{equation}
for brevity. Keeping the
full correlations we find the corrections for the first two emissions
to be given by
\begin{eqnarray}
w_{12}^2 = w_{21}^2 & = & 1\\\nonumber
w_{13}^3 = w_{23}^3 & = & \frac{9}{8}\\\nonumber
w_{31}^3 = w_{32}^3 & = & \frac{1}{2}\\\nonumber
w_{12}^3 = w_{21}^3 & = & -\frac{1}{8} \ .
\end{eqnarray}
The negative contribution from the $q\bar{q}$ dipole in the $q\bar{q}g$
system has already been noted in \cite{Azimov:1986sf}. Note that, {\it e.g.}
$w_{12}^3+w_{13}^3=1$ as dictated by color conservation.
In the shower approximation we have
\begin{eqnarray}
w_{12}^2 = w_{21}^2 & = & 1\\\nonumber
w_{13}^3 = w_{23}^3 & = & 1\\\nonumber
w_{31}^3 = w_{32}^3 & = & \frac{1}{2}\\\nonumber
w_{12}^3 = w_{21}^3  & = & 0
\end{eqnarray}
matching precisely the naive expectations on the subleading $\Nc$
contributions, that there is no radiation off the $q\bar{q}$ dipole
in a $q\bar{q}g$ system. For the first emission there is no difference
between the two approximations reflecting the triviality of the color
basis in that case. Also, gluon splittings in a $q\bar{q}g$ system
do not exhibit any subleading $\Nc$ correction.

More non-trivial dynamics are present for the fourth emission.
Introducing the relative magnitudes of dipole kernels encountered in
the four parton system with respect to the $q(\bar{q})g$ dipoles,
\begin{equation}
r_{ik} = \frac{V_{ik}^3+V_{ki}^3}{V_{13}^3+V_{31}^3+V_{23}^3+V_{32}^3}
\end{equation}
we find for the exact correlations
\begin{eqnarray}
w_{13}^4=w_{24}^4&=&\frac{9}{8}\ \frac{r_{23} - \frac{1}{9}r_{12}}{
1 - \frac{1}{9}r_{12}}
\\\nonumber
w_{14}^4=w_{23}^4&=&\frac{9}{8}\ \frac{r_{13} - \frac{1}{9}r_{12}}{
1 - \frac{1}{9}r_{12}}
\\\nonumber
w_{31}^4=w_{42}^4&=& \frac{1}{2}\ \frac{r_{23} - \frac{1}{9}r_{12}}{
1 - \frac{1}{9}r_{12}}
\\\nonumber
w_{41}^4=w_{32}^4&=& \frac{1}{2}\ \frac{r_{13} - \frac{1}{9}r_{12}}{
1 - \frac{1}{9}r_{12}}
\\\nonumber
w_{34}^4=w_{43}^4&=&
\frac{1}{2}\ \frac{1}{1 - \frac{1}{9}r_{12}}
\\\nonumber
w_{12}^4 = w_{21}^4 &=& -\frac{1}{8}\frac{1-\frac{10}{9}r_{12}}{1-\frac{1}{9}r_{12}} \ .
\end{eqnarray}
We note that the correction weights only depend on the quantities
$r_{ij}$. Conversely, in the large $\Nc$ limit we find
\begin{eqnarray}
w_{13}^4=w_{24}^4&=&r_{23}
\\\nonumber
w_{14}^4=w_{23}^4&=&r_{13}
\\\nonumber
w_{31}^4=w_{42}^4&=& \frac{1}{2}\ r_{23}
\\\nonumber
w_{41}^4=w_{32}^4&=& \frac{1}{2}\ r_{13}
\\\nonumber
w_{34}^4=w_{43}^4&=&\frac{1}{2}
\\\nonumber
w_{12}^4 = w_{21}^4 &=& 0 \ .
\end{eqnarray}
Again, there is no radiation off the $q\bar{q}$ dipole, the $gg$
dipole precisely matches the standard shower implementation.  The
different combinations of $q(\bar{q})g$ dipoles resemble the standard
shower implementation only in the case that -- in the three parton
system -- the splitting functions of one dipole had been much larger
than the splitting functions of the other dipole. In this case the
weights precisely match up the distribution of radiation generated by
the shower once it has decided which dipole was to radiate -- this
history is of course closely linked to the hierarchy of splitting
kernel values encountered. Indeed, the deviations between the shower
approximation and the standard shower implementation have been found
to be negligible.

The shower is terminated when either the $\Ng$:th gluon (we here
consider up to $\Ng=4,5,6$ gluon emissions) is emitted, or
when the infrared cut-off, taken to be $1\ {\rm GeV}$, is reached.
Only about $1 \%$ of all events radiate up to 6 gluons.
\FIGURE[b]{ \includegraphics[scale=0.7]{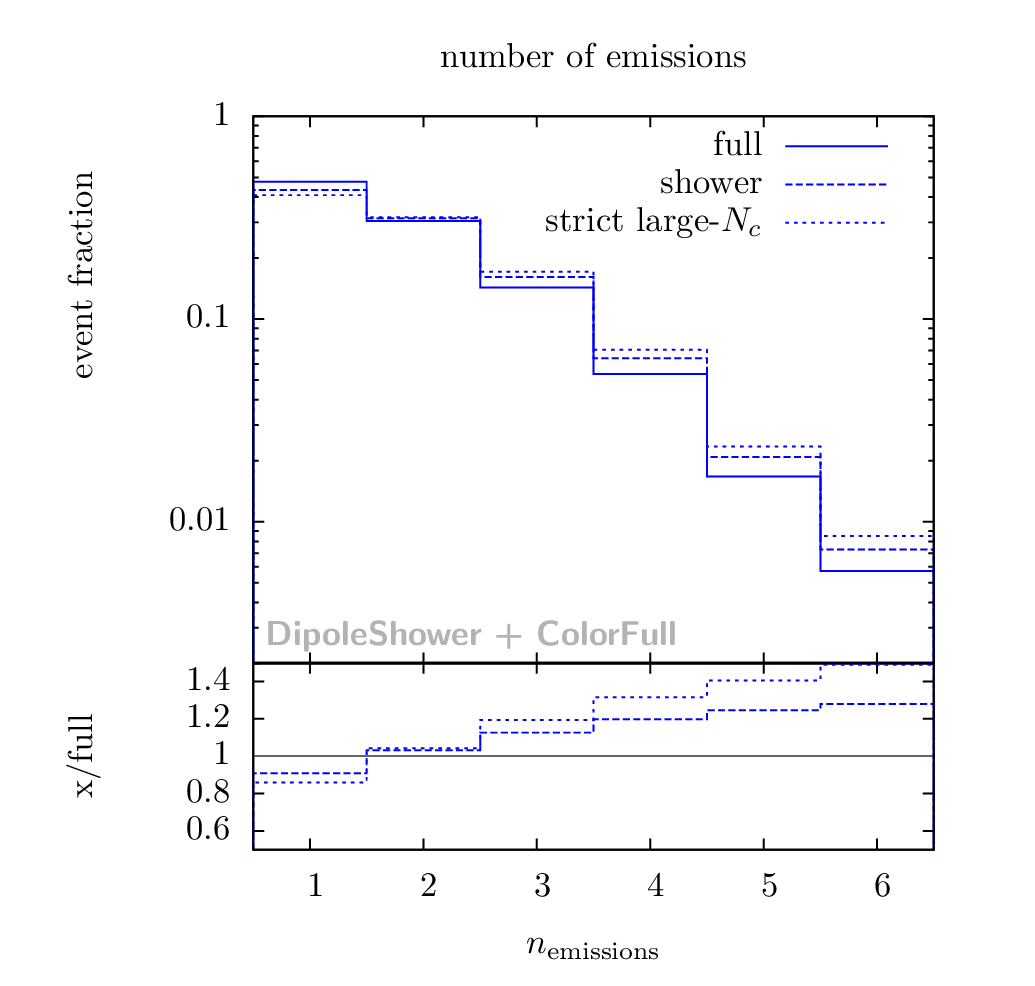}
  \caption{\label{fig:NEmissions} The normalized distribution of the number of
    emissions above the infrared cutoff comparing the different levels
    of approximation.} }
As, due to computational limitations, we only consider a limited
number of emissions, we have, for the observables considered here, 
checked that the prediction stabilizes when going up to six emissions. The
distribution of the number of emissions is shown in
\figref{fig:NEmissions} for the various approximations considered.
We find that taking the strict large $\Nc$ limit rises 
the probability for many gluon emissions by more than 40$\%$.
This can in part be attributed to the fact that $\cf=\TR (\Nc^2-1)/\Nc$, 
which equals $4/3$ (using $\TR=1/2$) is replaced by its leading part
$3/2$ also for collinear emissions. The buildup of this increased radiation 
probability accounts entirely for the difference between the 
"strict large $\Nc$" and the "shower" treatments. 
Apart from this effect, there is another effect 
which makes both the strict large $\Nc$ and the shower treatments 
of color, result in more radiation, compared to keeping the full color 
structure. 
This may be attributed to the fact that the
interference between color structures, where two gluons cross each other,
as depicted in in \figref{fig:1gCross}(b),
comes with negative sign. These terms
-- which are not present in standard parton showers -- thus
seem to lower the probability for radiation.
\FIGURE[t]{
  \includegraphics{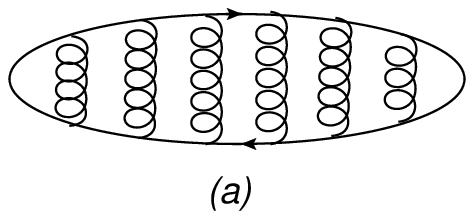}
  \includegraphics{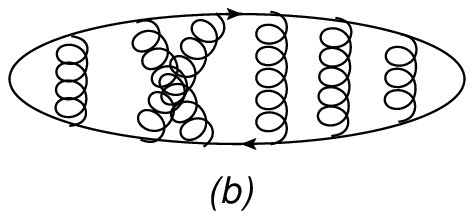}
  \includegraphics{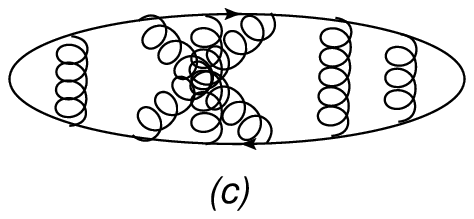}
  \caption{\label{fig:1gCross} With diagrammatic techniques an amplitude
    square, as in (a) can easily be evaluated to
    $\Nc \cf^{\Ng}$. The type of interference depicted in (b),
    where 2 gluons cross at one point contributes negatively
    to the radiation probability with $-\TR \cf^{\Ng-1}$, whereas
    the interference depicted in (c) contributes
    positively with $\TR^2 \cf^{\Ng-2}(\Nc^2+1)/N$ .
}}
For a large number of emissions, color structures where gluons cross each 
other in more complicated ways dominate the color suppressed 
contributions. 
Starting from two $\qqbar$-pairs there are terms which are only 
suppressed  by one power of $\Nc$.
We thus caution that the effects of color suppressed terms may be 
significantly larger at the LHC.

For the LEP-like setting considered here, we have investigated a set
of standard observables, event shapes and jet rates.  In all cases we
find the deviations of the shower approximation to be small, up to a
few percent, when compared to the full color treatment. The
differences between the strict large $\Nc$ and the "shower" treatment
option are often larger, in the cases of thrust, $T$, and sphericity up to
roughly 10$\%$.  In \figref{fig:OneMinusThrust-plot} we show the 
$\tau=1-T$ distributions for the three options of color structure
(left), and -- using the full color structure -- depending on how many
emissions we allow (right). The right plot shows that the prediction
stabilizes from five emissions onward.  \FIGURE[t]{
  \includegraphics[scale=0.7]{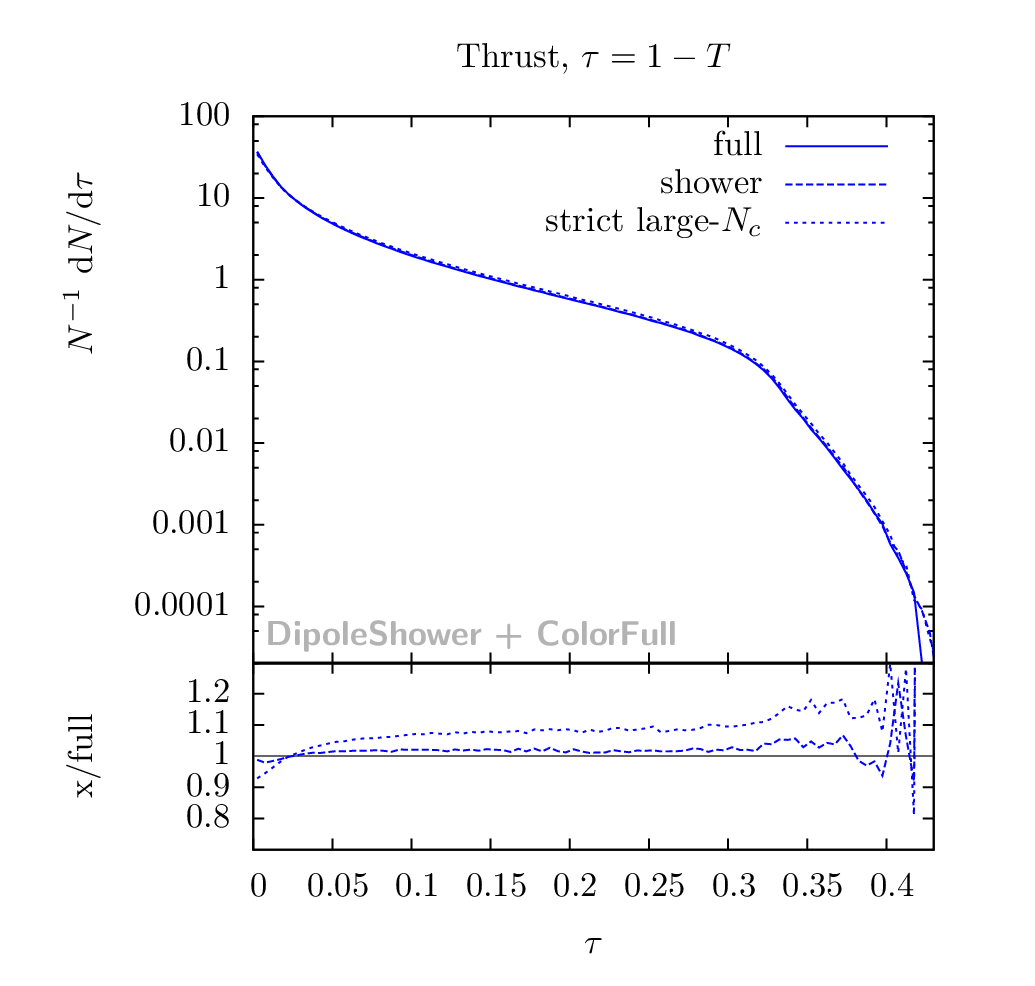}
  \includegraphics[scale=0.7]{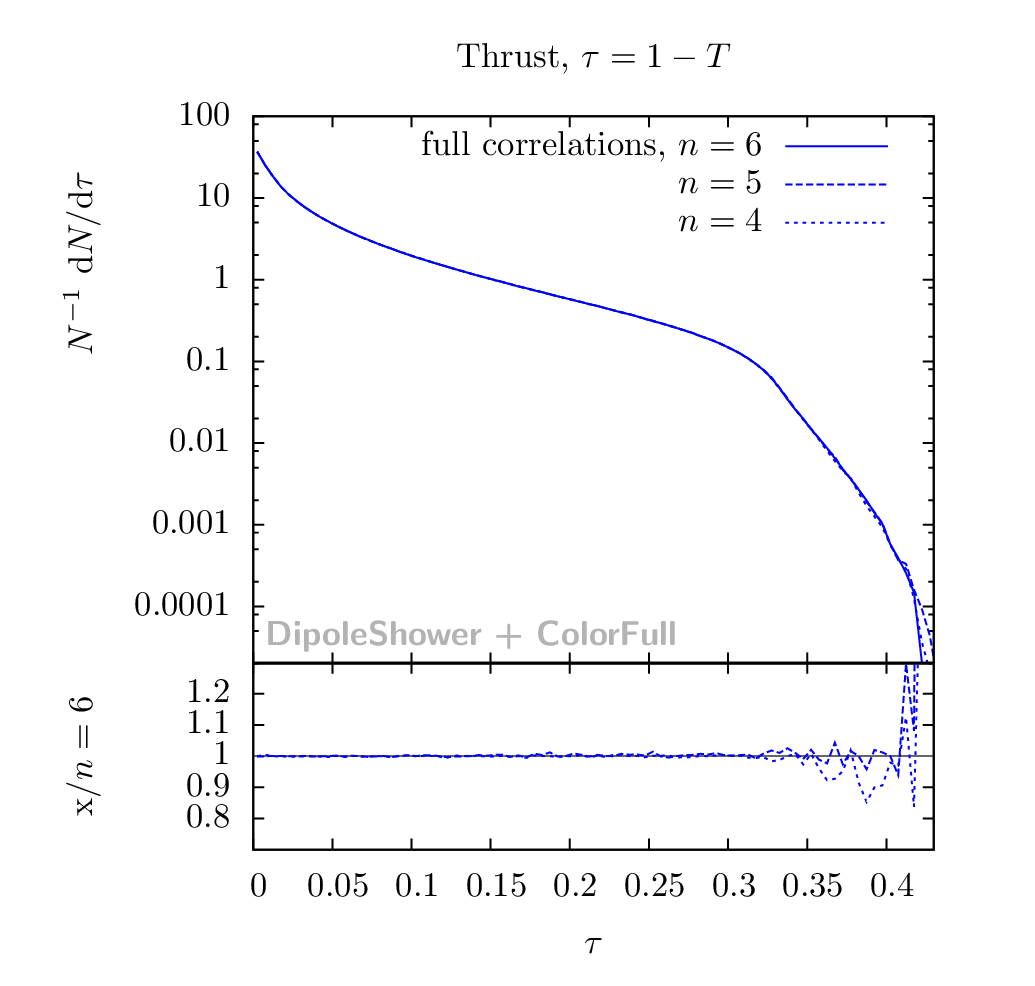}
  \caption{\label{fig:OneMinusThrust-plot} The distribution of one
    minus thrust, $\tau=1-T$. The left panel compares different
    approximations of the color structure. The right panel compares
    predictions depending on the number of emissions using the full
    color structure.  The prediction stabilizes from five emissions
    onward, as is the case for the other observables considered.}}
We note that there is not much difference between the full and the 
shower-like treatment of the color structure, whereas the strict large $\Nc$
treatment results in less pencil like events. This can be understood by
noting that there are more emissions in this case,
as indicated in \figref{fig:NEmissions}.

We now turn our attention to differential Durham $n$-jet rates,
\figref{fig:y_2_3-plot}, showing the distribution of the resolutions scale, 
$y=2\,\mbox{min}(E_i^2, E_j^2)(1-\cos{\theta_{ij}})/s$, at which an
$n+1$-jet event changes to an $n$-jet event.
We note that for the two-jet rate, there is
almost no difference between the full and "shower" option whereas,
increasing the number of jets, we may see a few percent difference,
although this difference may be just a statistical fluctuation.  
For the two jet rate the small difference
can be understood by considering that the transition from a three to a
two-jet event is {\it only} sensitive to the emission of a gluon from a
quark-line, i.e. to $\cf$, which is correctly described in standard
showers, as well as for the case that no non-trivial color
correlations are present for the first emission.
Nevertheless, the difference between shower
and full remains very small, at most a few percent, for the
differential $n$-jet observables.

\FIGURE[t]{
  \includegraphics[scale=0.7]{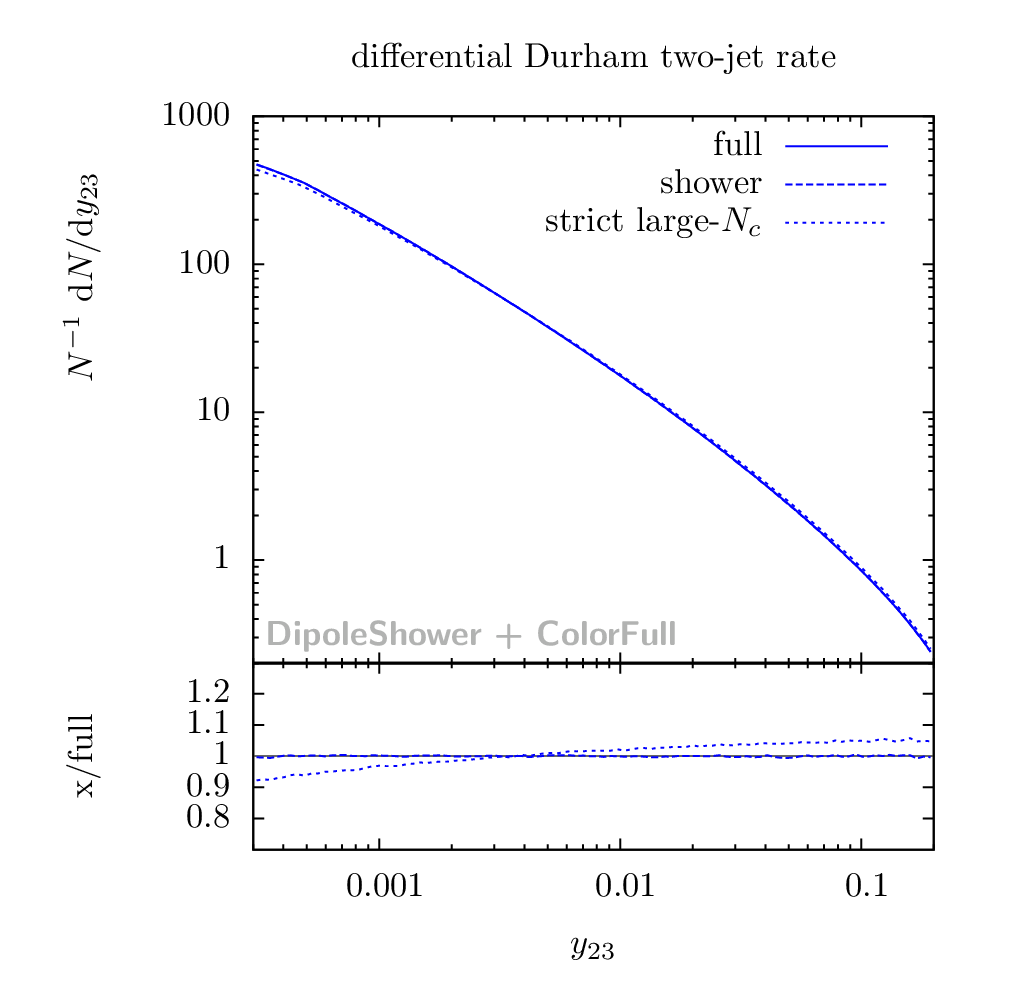}
  \includegraphics[scale=0.7]{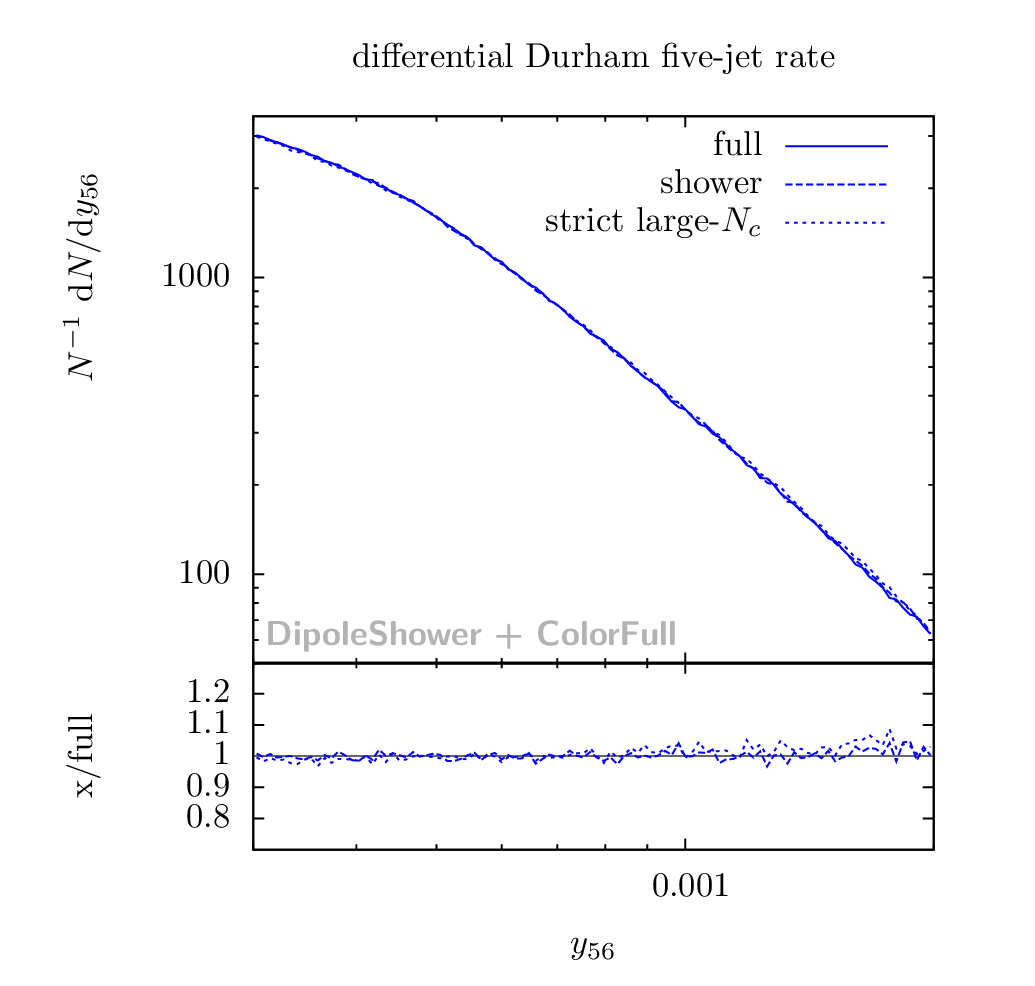}
  \caption{\label{fig:y_2_3-plot} 
Durham differential $n$-jet rates for transition from three to two
(left), and six to five jets (right).}}

In general we find that the differences between the shower approximation and 
full
correlations are small for event shapes and jet rates considered here.
The explanation for this is likely that these observables either are 
mainly sensitive to the collinear singularity -- which is treated with
the correct color factor in standard parton showers -- or are mainly
sensitive to the first hard emission, in case of which there are no
non-trivial correlations present.

The next candidates to check for larger effects are four-jet
correlations, which have been studied at LEP, mainly to investigate the
non-abelian nature of QCD, {\it i.e.} these are all very sensitive to
$g\to gg$ splittings. An example, the distribution of the cosine of
the angle between the softest two jets in four-jet events at a
Durham-jet resolution of $y=0.008$ are shown in
\figref{fig:cosa34-plot}. No large deviations are observed between the
different approximations.  A closer consideration of the color space
for a $q\bar{q}$ pair and two gluons reveals that this may actually be
expected.
Note that there is almost no difference between the shower
and strict large $\Nc$ approximations, which can be attributed
to the fact that these observables mainly probe gluon splitting which
is not sensitive to the difference between these approximations.
\FIGURE[b]{ \includegraphics[scale=0.7]{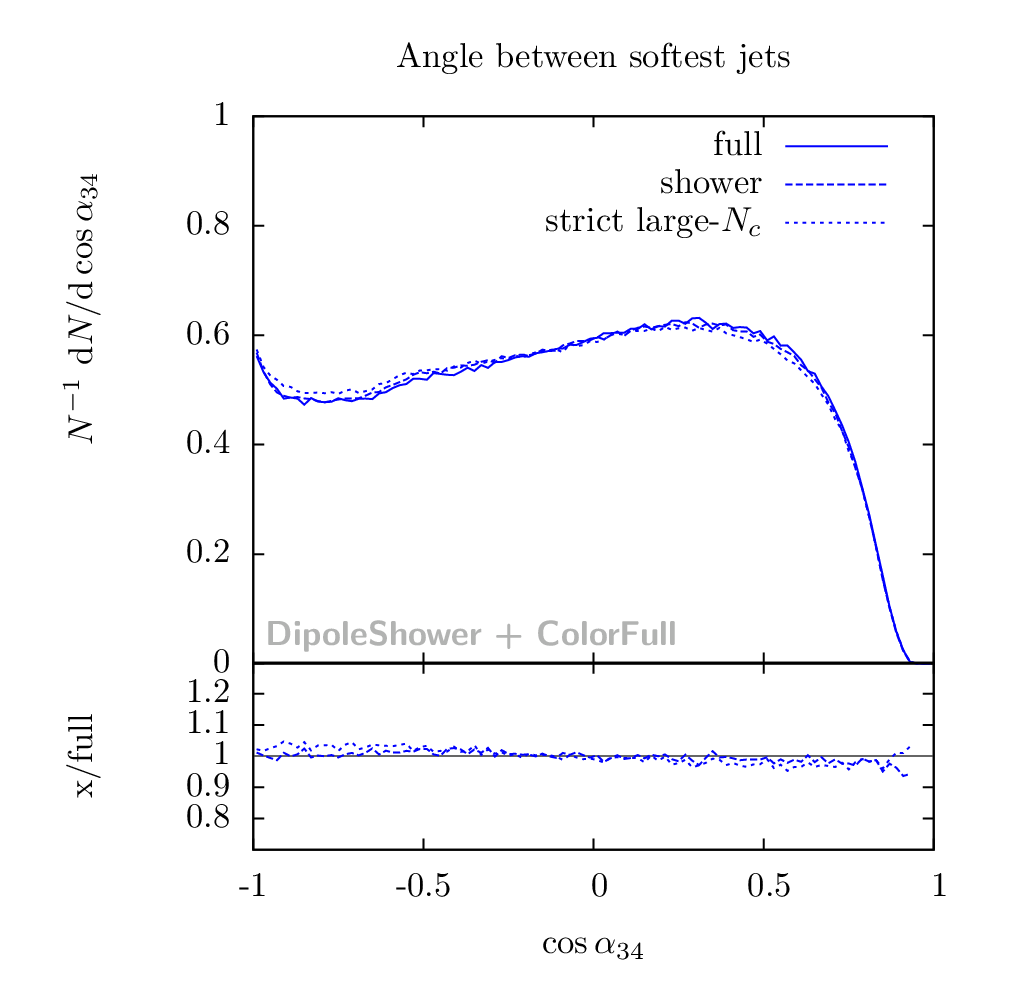}
  \caption{\label{fig:cosa34-plot} Distribution of the cosine of the
    angle between the softest jets in four-jet events.}}

To investigate the effects of soft coherent emissions, we have
therefore also studied the average transverse momentum and rapidity of
parton four onward with respect to the thrust axis defined by the
three hardest partons.  The result is shown in
\figref{fig:avgpt-plot}.  \FIGURE[t]{
  \includegraphics[scale=0.7]{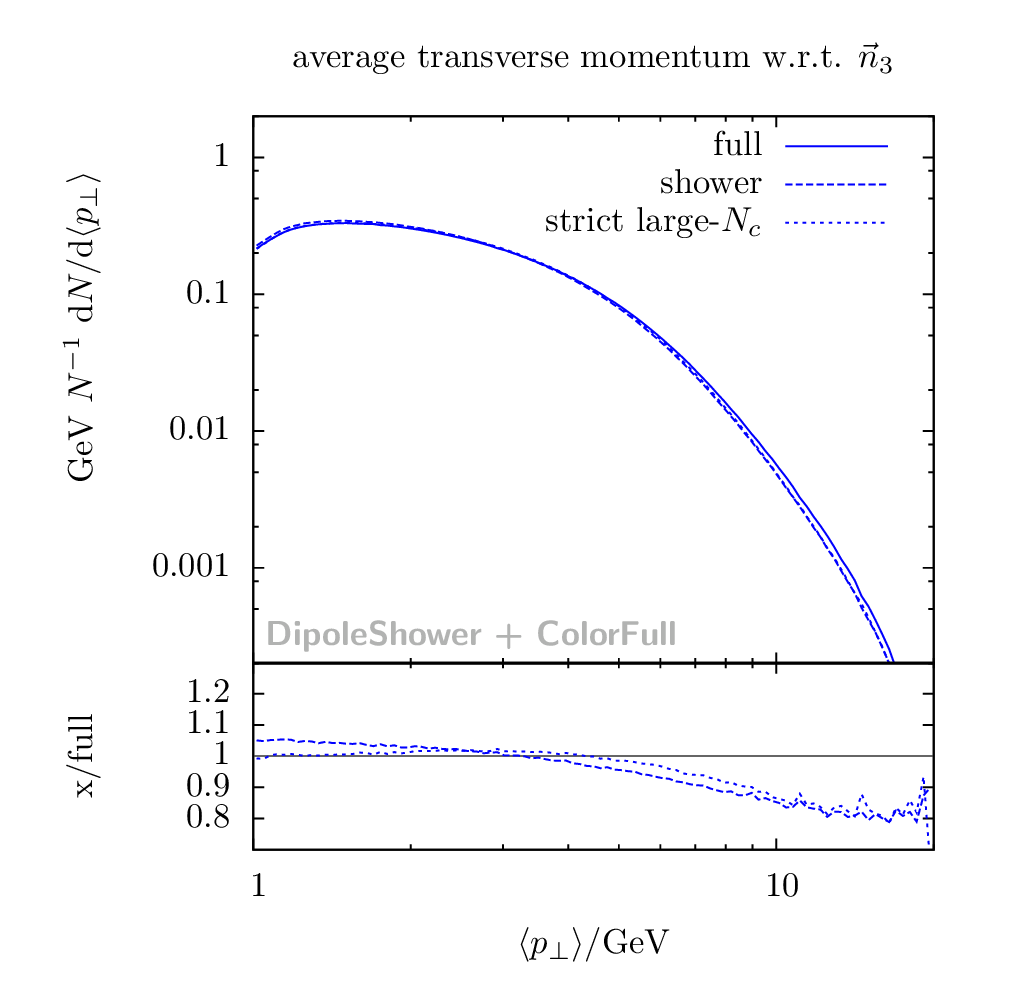}
  \includegraphics[scale=0.7]{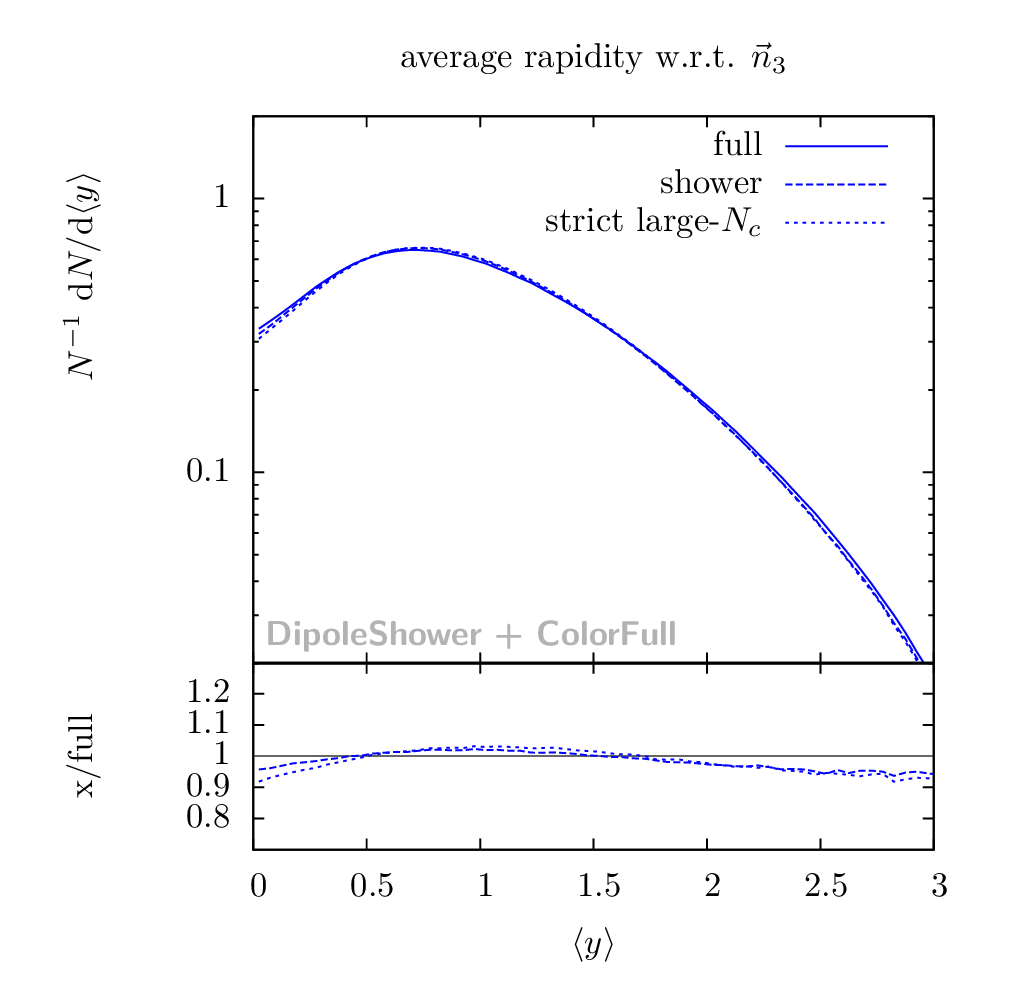}
  \caption{\label{fig:avgpt-plot} The average transverse momentum and rapidity of
    partons four onward w.r.t. the thrust axis defined by the three hardest
    partons.}}  Here the effects are much larger.  We find that the
average transverse momentum is harder if the full color structure is
kept, for large $\langle p_\perp\rangle$ up to $20 \%$. 
We also see that the strict large $\Nc$ approximation is
somewhat closer to the prediction including the full
correlations. We leave a
more detailed study of this class of observables as future work,
including an operational definition in terms of anti-$k_\perp$-type
jets to build up a reference system from hard, collinear jets while
being able to look at the orientation of soft radiation relative to
this system.

\section{Conclusions and Outlook}
\label{sections:conclusions}

In this paper the first results from a subleading $\Nc$ parton shower were presented. 
We have considered final state gluon emissions using iterative
matrix element corrections to treat the full SU(3) structure for each
subsequent emission. 

A major conclusion is that standard LEP-observables, including event
shapes and jet rates, are only affected by at most a few percent.  For
a class of observables dividing the event into a hard reference system
and accompanying radiation we expect larger differences. This is
indeed seen for tailored observables like the average transverse
momentum with respect to a thrust axis determined by a system of three
hardest partons, in case of which we see deviations as large as $20\%$.

The small differences can, in the case of the standard observables,
largely be explained by the fact that these observables are either
sensitive to collinear radiation, or to the first hard
emissions. 

Another contributing factor is that the color suppressed
terms can be seen to be quite small from considerations of color space alone, 
when starting from a $\qqbar$-pair. This is not the case if the hard 
scattering process is e.g. QCD $2 \to 2$.
In this case we may see more striking differences between the
approximations considered, though we cannot yet make a definite
statement. The simulation framework is general enough to cope with
this case and we leave a detailed discussion of subleading $\Nc$
effects at hadron colliders to future work.

In addition to what is presented here, there are several other effects
which should be included before it can be claimed that a parton shower
fully simulates SU(3) physics. 
Apart from including a running coupling constant and the 
$g \rightarrow \qqbar$ splitting kernel, we like to include virtual color 
rearranging gluon exchanges.
We view this work as a proof of concept, and a first step
towards quantifying the impact of subleading $\Nc$ contributions.

Finally, the parton shower outcome should be hadronized. While
this is in itself an interesting task, it is much beyond the scope of 
the present paper. In an ordinary parton shower, where the shower outcome  
corresponds to a well defined probabilistic color line arrangement,
this color structure is fed into the hadronization model.
Here, the state after showering contains amplitude level information,
and can thus not simply be input into existing hadronization models.
While studying the influence of the hadronization on the parton
shower outcome is an interesting task, we thus refrain from it at this
stage.

\section*{Acknowledgments}

We are grateful to Yuri Dokshitzer, Stefan Gieseke, Leif L\"onnblad
and Zoltan Nagy for discussions on this subject. We would also like to
thank Markus Diehl, Stefan Gieseke, G\"osta Gustafson and Torbj\"orn
Sj\"ostrand for comments on the manuscript.  S.P. acknowledges the
kind hospitality of the Theoretical High Energy Physics group at Lund
where this work was completed, and funding by the Helmholtz Alliance
``Physics at the Terascale''. M.S. was supported by the Swedish
Research Council (contract number 621-2010-3326).

\bibliography{colorshower}

\end{document}